\documentclass[prl,superscriptaddress,floatfix,showpacs,aps,reprint,letterpaper,nobalancelastpage]{revtex4-1}

\usepackage{amsmath}
\usepackage{graphicx}
\usepackage{xcolor}

\usepackage[urlcolor=blue, hyperindex, colorlinks, bookmarks=true]{hyperref}

\newcommand{\avg}[1]{{\langle{#1}\rangle}}

\newcommand{\bra}[1]{\langle #1|}
\newcommand{\ket}[1]{|#1\rangle}
\newcommand{\ketbra}[2]{|{#1}\rangle\!\langle{#2}|}
\newcommand{\var}[1]{\mathrm{var}(#1)}
\newcommand{\pd}[1]{\frac{\partial}{\partial #1}}

\usepackage{microtype}

\begin{document}

\title{Tomography via Correlation of Noisy Measurement Records}
\date{\today}

\author{Colm A. Ryan}
\author{Blake R. Johnson}
\affiliation{Raytheon BBN Technologies, Cambridge, MA 02138, USA}
\author{Jay M. Gambetta}
\author{Jerry M. Chow}
\affiliation{IBM T.J. Watson Research Center, Yorktown Heights, NY 10598, USA}
\author{Marcus P. da Silva}
\affiliation{Raytheon BBN Technologies, Cambridge, MA 02138, USA}
\author{Oliver E. Dial}
\affiliation{IBM T.J. Watson Research Center, Yorktown Heights, NY 10598, USA}
\author{Thomas A. Ohki}
\affiliation{Raytheon BBN Technologies, Cambridge, MA 02138, USA}

\begin{abstract}

We present methods and results of shot-by-shot correlation of noisy
measurements to extract entangled state and process tomography in a
superconducting qubit architecture. We show that averaging continuous values,
rather than counting discrete thresholded values, is a valid tomographic
strategy and is in fact the better choice in the low signal-to-noise regime. We show that
the effort to measure $N$-body correlations from individual measurements
scales exponentially with $N$, but with sufficient signal-to-noise the
approach remains viable for few-body correlations. We provide a new protocol
to optimally account for the transient behavior of pulsed measurements.
Despite single-shot measurement fidelity that is less than perfect, we
demonstrate appropriate processing to extract and verify entangled states and
processes.

\end{abstract}

\maketitle

By engineering coherent manipulation of quantum states we hope to gain
computational power not available to classical computers. The purported
advantage of a quantum information processing system comes from its ability to
create and control quantum correlations or entanglement \cite{Jozsa2003}. As we
work towards obtaining high-fidelity quantum control, the ability to measure,
confirm and benchmark quantum correlations is an important tool for debugging
and verification.

Measurement results for only parts of a system by themselves are insufficient
to reconstruct the overall state of the system, or of a process acting on the
system.  These sub-system statistics cannot capture correlations, whether
classical or quantum. Joint quantum measurements, ones that indicate the
joint quantum state of a system, can provide information about the entire
system, and thus allow for complete reconstruction of states or processes.
These joint measurements may be intrinsically available or they may be
effectively enabled by post-measurement correlation of the single-qubit
results.

The physics of particular systems may enable such intrinsic joint
measurements---for example, in a circuit quantum electrodynamics (QED) setup,
the frequency shift of a resonator coupled dispersively to multiple qubits
provides information about the joint state of the qubits
\cite{Chow2010,Filipp2009}, or in an ion-trap system the total fluoresence
from a chain of ions provides similar joint information \cite{Leibfried2005}.
However, there is a limit to the correlations that can be engineered while
maintaining individual addressability and control. Alternatively,
entangling operations between qubits, before their individual measurement,
can implement effective joint measurements by mapping a multi-body observable
to a single-qubit one \cite{Liu2005a}, but these operations can also be error-prone or
difficult to implement. Instead, an accessible approach to joint measurements
is to measure individual qubits and {\em correlate} the separate single-shot
records. This approach is common to other solid state qubits
\cite{McDermott2005,Nowack2011}, optical photonics \cite{Altepeter2004} and microwave optics \cite{Bozyigit2010,Eichler2012}.

When constructing joint measurements from correlating individual ones, one
must address the requirements for the individual measurements in order to
obtain high-fidelity reconstruction of multi-qubit states and processes. In
particular, since errors in the individual measurements will propagate into
correlations, one might wonder if low-fidelity individual measurements make it
difficult or impossible to reconstruct entangled states. Fortunately, circuit
QED allows for high-\emph{visibility} measurements even when the single-shot
fidelity is poor \cite{Wallraff2005}, meaning that the dominant measurement
noise is state-independent. This allows for a strategy that is not available
with traditional counting detectors: correlate the \emph{continuous}
measurement response without thresholding into binary outcomes. We will show
that averaging such continuous outcomes allows for an unbiased estimate of
multi-qubit observables, and is the best strategy in the low signal-to-noise (SNR) regime.

Whenever correlating noisy individual measurements, there is reduced SNR in
the correlated measurements. We will show that for $N$-body correlations this
SNR penalty is generally exponential in $N$, but that in the large SNR regime
reduces to $1/N$. This implies that a greater amount of averaging is necessary
to obtain accurate estimates of multi-body terms than single-body ones, but
for the SNR available in current experiments, few-body correlations remain
readily accessible.

Certain quantum computing architectures, such as the surface code \cite{DiVincenzo2009},
require individual qubit measurements which must be correlated for debugging
and validation purposes. In a subsection of a circuit QED implementation of such
an architecture, we present a phase-stable heterodyne measurement technique and a
new filter protocol to optimize the SNR of a pulsed measurement. Using these
techniques we verified that our measurements have a highly single-qubit
character. Finally, we demonstrate the viability of correlating noisy
measurement records by characterizing two-qubit states and entangling
processes.

\section{Soft-averaging vs. Thresholding}

Measurement in a circuit QED setup typically consists of coupling the qubit to
an auxiliary meter (usually a cavity mode), and then inferring the qubit state
from directly measuring the auxiliary mode. Since the qubit-mode coupling is
effectively diagonal in the qubit's eigenbasis, the qubit POVMs corresponding
to different measurement outcomes will always be diagonal as well, even when
the measurement is not projective (i.e., when there is a finite probability of
error for the measurement to distinguish between excited and ground states).
Although the measurement of the meter can in principle take an unbounded
continum of values, discrete outcomes can be obtained by {\em thresholding} to
bin those measurement outcomes into a finite set. However, for tomography we
are more interested in estimating the expectation value of observables than in
the single-shot distinguishability of states. In this case, {\em soft-averaging}
of the measurement records over many shots, without thresholding,
can yield significant advantages.  

A simple model of circuit QED measurement illustrates this quite vividly. In
the absence of relaxation, measurements of the meter will yield a weighted
mixture of Gaussian distributions. Each of these Gaussian distributions will
have a mean and variance determined by details of the device, and the weights
correspond to the probabilities of the qubit being in different eigenstates.
In other words, conditioned on the qubit being prepared in an eigenstate $i$,
the distribution of measurement outcomes is $N(\mu_i,\nu_i^2)$, where $\mu_i$
is the measurement eigenvalue. Soft-averaging over $R$ shots scales the
variance of the distributions by a factor of ${1\over R}$, but an estimate of
any diagonal observable will have an unbiased mean. Thresholding, on the other
hand, will result in biased estimates of the expectation value because for
Gaussian distributions there is always a finite probability of mistaking one
measurement outcome for another. This bias can be corrected by rescaling the
measurement results which converts the bias into additional variance. Soft-averaging
also requires scaling to translate from measurement eigenvalues to
the $\pm1$ outcomes expected for a Pauli operator measurement  \footnote{In
both cases this can lead to systematic errors if coherent imperfections are
present, e.g. wrong basis from rotation errors. Full measurement tomography,
with perfect state preparation, could correct for these more general biases.
Or, in the case of soft-averaging access to a calibration that is state-preparation
independent could also side-step this issue.}. The mean-squared
error (MSE), equal to the sum of the variance and the square of the bias, is a
figure of merit one can use to evaluate how well we can estimate an
observable. If we assume the bias can be correted then the best strategy
is determined by the relative variance of the two approaches which will depend on
the SNR of the individual measurements and the number of averages.  In the low
SNR regime soft-averaging is preferred whereas in the high SNR regime
thresholding is better.

\begin{figure}[tb]
\includegraphics[width=\columnwidth]{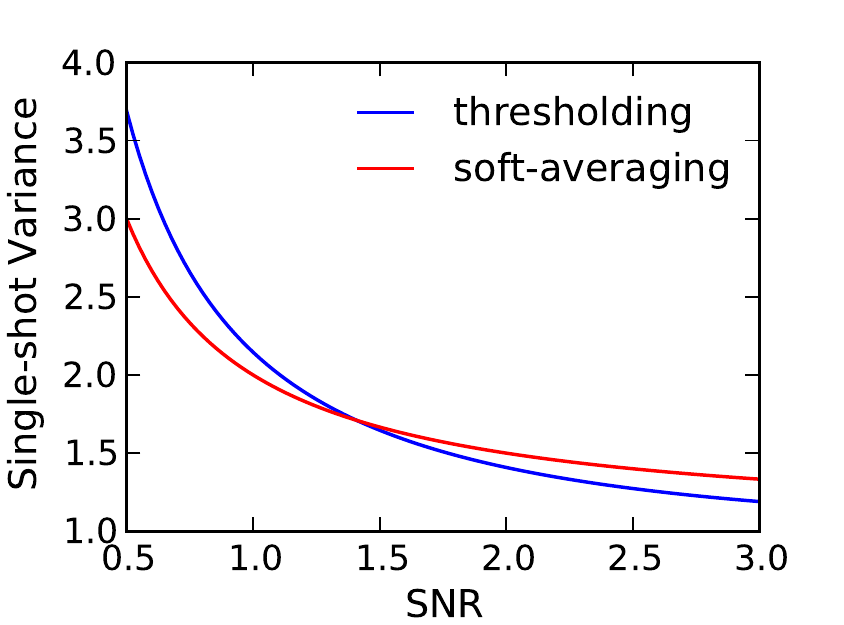}
\caption{\label{variance-crossover} 
The variance of soft-averaging and thresholding as a function of measurement
SNR for a superposition state ($\langle\sigma_z\rangle=0$). At low SNR soft-averaging
is the better choice but at high SNR it pays an extra penalty from the widths
of the distributions whereas thresholding is only affected by the exponentially
small overlap of the two Gaussians.}
\end{figure}

As a simple concrete example, consider the measurement of a single qubit,
where $\nu_0=\nu_1=\nu$ and $\mu_0=-\mu_1=1$. For an arbitrary state, soft-averaged
estimates of $\avg{\sigma_z}$ will be distributed according to
$N(\avg{\sigma_z},{\nu^2 \over R}+{1-\avg{\sigma_z}^2 \over R})$, where the
first variance term is the instrinic Gaussian variance and the second is
quantum shot noise. Setting the threshold at zero, the thresholded estimates
will have a mean given by $\avg{\sigma_z} [2\Phi(1/\nu\sqrt{2}) - 1]$, where
$\Phi$ is the cummulative distribution function of a normal random variable,
and variance of ${1\over R}-{\avg{\sigma_z}^2\over R} [2\Phi(1/\nu\sqrt{2}) - 1]^2$.
Consequently, thresholding will introduce a bias of $2 \avg{\sigma_z}
[1-\Phi(1/\nu\sqrt{2})]$, which is independent of the number of averages, $R$.
If we now assume we have perfect knowledge of the bias from calibration
experiments then the rescaled thresholded variance is
\begin{equation}
{1\over R [ 2 \Phi(1/\nu\sqrt{2}) - 1]^2} - {\avg{\sigma_z}^2\over R},
\end{equation}
whereas it has not changed for soft-averaging. The $\mathrm{SNR} =
1/\nu^2$ where the variances are equal occurs at 
\begin{equation}
{1\over [ 2 \Phi(1/\nu\sqrt{2}) - 1]^2} = \nu^2 + 1,
\end{equation}
which is satisfied at $\mathrm{SNR} \approx 1.41$ (corresponding to a single-shot
fidelity of 76\%) and is unchanged as one correlates multiple
measurements.  Above this cross-over, soft-averaging pays an additional
variance penalty which is exponentially worse than thresholding; however, as
seen in Figure \ref{variance-crossover}, in this regime we are often limited
by quantum shot noise. Substantial advances in measurement fidelity in
superconducting circuits due to the use of quantum-limited amplifiers\cite
{Castellanos-Beltran2008, Bergeal2010, Hatridge2011}, implies that the
community will shortly enter the regime where thresholding is the preferred
strategy.

\section{SNR Scaling of Correlation Terms}

There is a cost associated with using the correlations of subsystem
measurements instead of joint measurements, which can be most easily
illustrated by seeing how the accuracy of measurement estimates scales with
the variance of the observations~\cite{DaSilva2010}.  Consider a product state
such that the measurement records $X_i$ are independent random variables with
mean and variance $(\avg{\sigma_{z,i}},\nu_{i}^2 + (1-\avg{\sigma_{z,i}}^2))$. Then the variance of the correlated records of $N$ subsystems is given by \cite{Goodman1962},
\begin{align}
\label{eq:bare-corr} 
\nu_{\mathrm{corr}}^2 &= {\mathrm{Var}}(X_{i_1}X_{i_2}\cdots X_{i_N}) \nonumber \\
&= \Pi_{k=1}^N (\nu_{k}^2 + (1-\avg{\sigma_{z,i}}^2) + \avg{\sigma_{z,i}}^2) - \Pi_{k=1}^N \avg{\sigma_{z,i}}^2 \nonumber\\
& = \left[\Pi_{k=1}^N \left(\nu_{k}^2 + 1\right) - \avg{\sigma_{z,1}\cdots\sigma_{z,N}}^2\right]. 
\end{align}
$\avg{X_{i_1}X_{i_2}\cdots X_{i_N}}^2$ is an unbiased estimate for
$\avg{\sigma_{z,1}\cdots\sigma_{z,N}}^2$ so the MSE is
equal to $\nu_{\mathrm{corr}}^2$ and it grows exponentially with the number of
correlated terms. Entangled states will have correlated shot noise and the variance calculation is considerably more involved.  However, for tomography of arbitrary states we are limited by this exponential scaling.  It is possible to reduce the variance by repeating the
measurement $R$ times and averaging, but in order to get some fixed accuracy
on the estimate of $\langle X_{i_1}X_{i_2}\cdots X_{i_N}\rangle$, $R$ will
still have to scale exponentially with $N$.

For small $N$, and equal and sufficiently high SNR ($\nu_{k}^2 = 1/\mathrm{SNR} \ll 1$), then Eq.~\ref{eq:bare-corr} reduces to, 
\begin{equation}
\nu_{\mathrm{corr}}^2 \approx \frac{N}{\mathrm{SNR}},
\end{equation}
so $R$ is simply linearly related to $N$ in this favorable regime. Thus, measurements of low-weight correlators are still accessible without a punative experimental effort.

\section{Experimental Setup}

\begin{figure}[tb]
\includegraphics[width=\columnwidth]{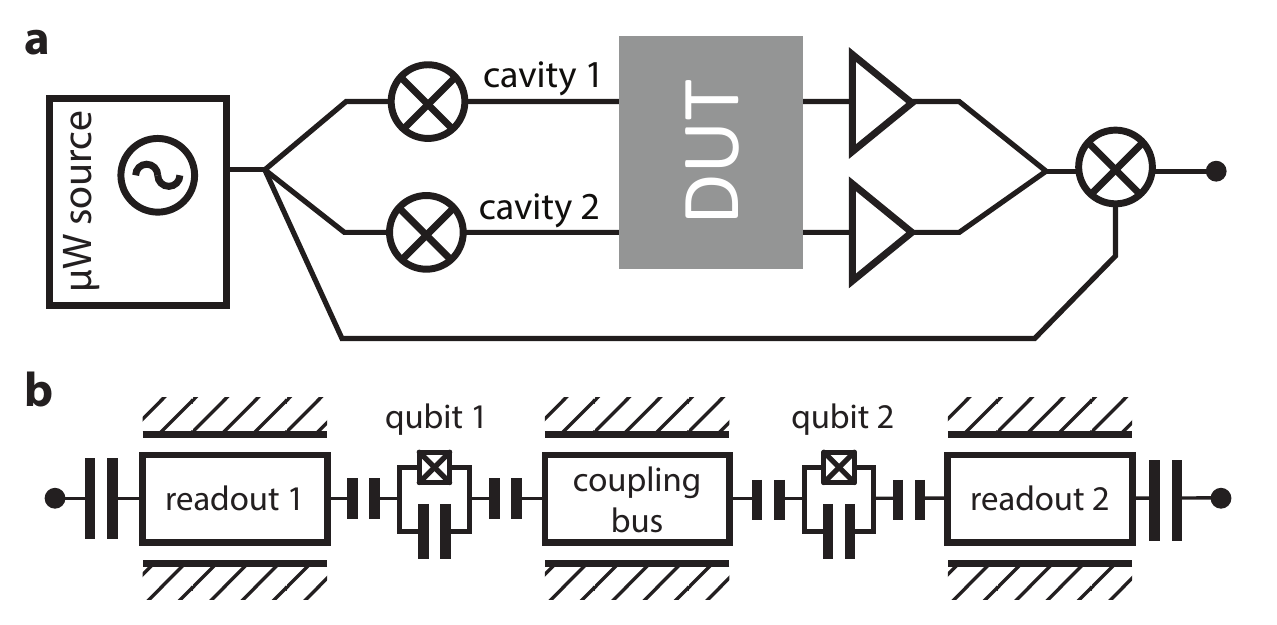}
\caption{\label{ExpSchematic} (a) Schematic for a parsimonious two
  channel autodyne measurment.  A single microwave source is split and
  both channels are SSB I/Q modulated to produce a shaped pulse at the
  cavity frequency. The refected signal is then amplified and mixed
  back down with an LO from the same microwave source.  (b) Schematic
  of the two-qubit system with individual readout resonators and an
  unconnected coupling bus resonator.}
\end{figure}

Samples were fabricated with three single-junction transmon qubits in linear
configuration with nearest-neighbours joined by bus resonators \cite{Chow2012}.
For the experiments discussed here, we used two of the three qubits, so
the relevant subsystem is as shown in Fig.~\ref{ExpSchematic}(b). Similar
chips were measured at IBM and BBN. Each qubit has an individual measurement
resonator coupled to it via the standard circuit QED Hamiltonian
\cite{Blais2004}. The same resonator is also used for driving the qubit
dynamics with microwave pulses. The resonant frequency of the cavity exhibits
a qubit-state dependence which we measure by the response of a microwave pulse
applied near the cavity frequency.

We employed an ``autodyne'' approach, similar to Ref. \cite{Jerger2012} and
shown in Fig.~\ref{ExpSchematic}(a), to measure the qubit state via the
reflection of a microwave pulse off the coupled cavity. Autodyning produces a
heterodyne signal from a single local oscillator (LO). The microwave LO,
detuned from the cavity, is single-sideband (SSB) modulated via an IQ mixer to
bring the microwave pulse on-resonance with the cavity. The reflected
amplified signal is then mixed down with the same microwave carrier.  This
eliminates the need for two microwave sources, nulls out any microwave phase
drifts and moves the measurement away from DC. If the SSB modulation comes
from an arbitrary waveform generator (AWG) it also allows for measurement
pulse shaping. In addition, if multiple read-out cavities are close in
frequency, it allows us to use a single microwave source to drive multiple
channels with relatively less expensive power splitters and amplifiers.

At the readout side, the ability to choose the heterodyne IF frequency allows
multiple readout channels to be frequency muliplexed onto the same high-speed
digitizer and then digitally separated using techniques from software defined
radio.  Although our current implementation is purely in software, we expect
it to readily transfer to hardware as we scale up the number of readout
channels \cite{mchugh:044702}.  The initial data stream is sampled at 500MS/s
and is immediately decimated with a low-pass finite-impulse response (FIR)
filter.  This allows us to achieve good phase precision with the relatively
low vertical resolution (8-bits) of our digitizer card (AlazarTech 9870).  The
channels are then extracted with a frequency shifting low-pass filter.  The
bandwidth needed per channel is $(2\chi + \kappa)/2\pi$, where $\kappa$ is the
cavity linewidth, and $\chi$ the dispersive shift. For current parameters,
this gives channels bandwidths of a few MHz. Future devices optimized for
high-fidelity readout will have larger $\chi$ and $\kappa$, increasing the
channel bandwidth to $\approx 10\,\mathrm{MHz}$. This is much smaller than the
typical analog bandwidth of commercial digitizing hardware, allowing many
readout signals to be multiplexed onto a single physical channel.

\section{Measurement Tomography} 

Before analyzing correlated sub-system measurements, it helps to first
optimize each individual readout and reduce the signal to a single quadrature.
In the absence of experimental bandwidth constraints, elegant solutions to the
optimal measurement filter exist \cite{Gambetta2007}. However, when the
measurement is made through a cavity, the measurement response is similar to
that of a kicked oscillator \cite{Gambetta2008} producing a phase transient in
the response. During the rising edge of the pulse the reflected signal's phase
swings wildly.  Unfortunately, this is exactly the most crucial time in the
record because it has been least affected by $T_1$ decay.  Simple integration
over a quadrature will lose information as the different phases cancel out.
Here we use the mean ground and excited state traces as a matched filter
\cite{Turin1960} to unroll the measurement traces and weight them according to
the time-varying SNR.  The matched filter is optimal for SNR, which would also
be optimal for measurement fidelity  in the absence of $T_1$. For the
experiments considered here, $T_1$ is much greater than the cavity rise time,
$1/\kappa$. In this case, the filter is close to the optimal linear filter.

To derive the matched filter consider that the measurement signal is given by
\begin{equation}
  \psi(t) \propto {1\over 2}[\alpha_0(t) - \alpha_1(t)]\sigma_z + \xi(t),
\end{equation}
where $\alpha_i(t)$ is the time-dependent cavity response when the qubit is in state $i$, and $\xi(t)$ is a zero-mean noise term that is uncorrelated with the state. The filtered measurement is given by an integration kernel, $K(t)$, such that
\begin{align}
  S &= \int_0^t K(t)\psi(t)\,\mathrm{d}t,\\
    &= \sum_j K_j \psi_j,
\end{align}
where in the second form we discretize time such that $K_j = K(t_j)$ and $\psi_j = \psi(t_j)$. From this we can see that
\begin{align}
  \avg{\Delta S} &= \sum_j K_j\avg{\alpha_0(t_j) - \alpha_1(t_j)},\\
  \nu^2 &= \var{\Delta S} = \sum_j K_j^2 \nu_j^2,
\end{align}
where $\avg{\Delta S}$ is the average difference of $S$ for $\sigma_z = \pm 1$ and $\nu_j^2 = \var{\alpha_0(t_j) - \alpha_1(t_j)} + \var{\xi}$. Then, to optimize the SNR we set $\pd{K_j}\left|\avg{S}\right|^2/\nu^2 = 0$. After dropping scaling factors that are independent of $j$, we find
\begin{equation}
  K_j = \frac{D^*(t_j)}{\nu_j^2},
\end{equation}
where $D(t_j) = \avg{\alpha_0(t_j) - \alpha_1(t_j)}$ is the difference vector
between the mean ground-state and excited-state responses. Since $T_1$
prevents fixing $\sigma_z$, we do not have direct access to $D(t)$.
Consequently, we approximate it by measuring the mean cavity response after
preparing the qubit in $\sigma_z$ basis states. Note that $\angle D$ gives the \emph
{time-dependent} quadrature containing qubit-state information. Thus, the
above construction rotates all information into the real part of the resulting
signal, so one can discard the orthogonal imaginary quadrature. We
additionally subtract the mean response to remove the identity component in
the measurement. This ensures that the resulting correlators are composed
mostly of multi-body terms. Finally, the optimal integration time is
determined by maximizing the single-shot fidelity.

\begin{figure}[tb]
\includegraphics[width=\columnwidth]{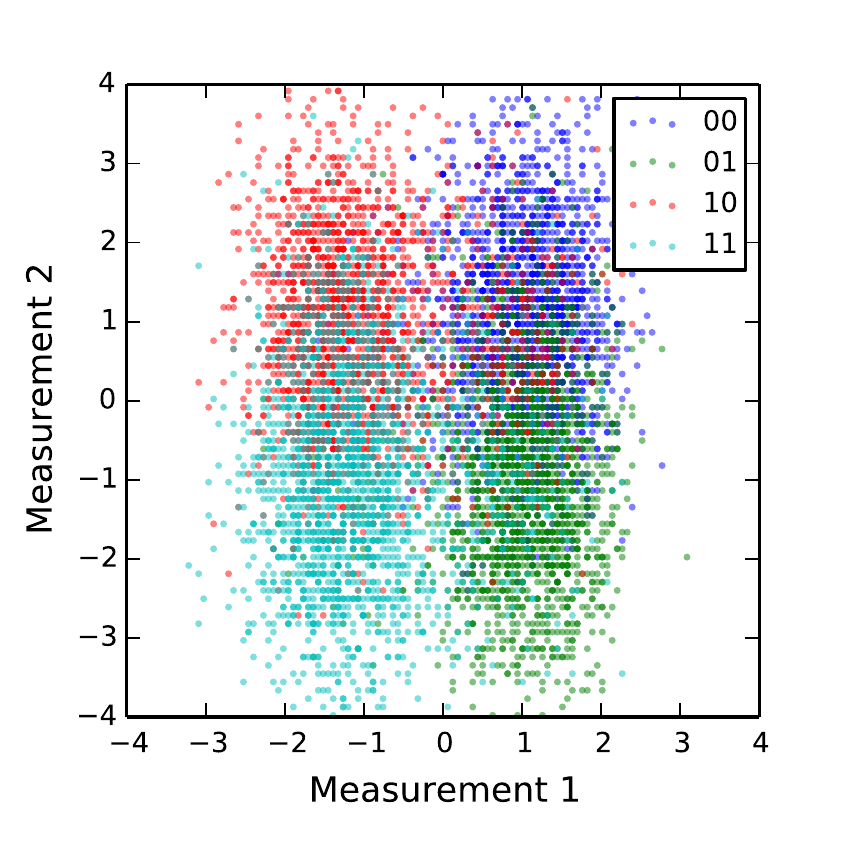}
\caption{\label{MeasScatter}
Scatter of the filtered measurement results for both readout
channels after preparation of the basis states $\ket{00}$ (blue), $\ket{01}$
(green), $\ket{10}$ (red), and $\ket{11}$ (cyan).  Qubit 2 has a
smaller effective coupling to its readout cavity, leading to increased
variance along the measurement 2 axis.}
\end{figure}

In the tight confines of the chip there is inevitable microwave coupling
between nominally independent lines that may inadvertently enable spurious
multi-qubit readout. Therefore, it is important to confirm that the
measurements give mostly independent single-qubit information and that our
joint readout is enabled only from post-measurement correlation of the
results. After verifying a sufficient level of qubit control with randomized
benchmarking \cite{Magesan2011} we run a limited tomography on the measurement
operators. By analyzing the measurements associated with preparing the four basis
states, shown in Fig.~\ref{MeasScatter}, we can convert to expectation values
of diagonal Pauli operators \cite{Chow2010} and confirm limited multi-Hamming
weight operators.
\begin{align}
\hat M_1 &= 1.0110(4)ZI + 0.0164(6)IZ - 0.0106(6)ZZ \nonumber \\
\hat M_2 &= 0.00(1)ZI + 0.98(1)IZ + 0.02(1)ZZ\\
\hat M_{1,2} &= 0.00(2)ZI + 0.00(2)IZ + 0.98(2)ZZ. \nonumber  
\end{align}

Given access to the single-shot measurement records, it is possible to
estimate the variances associated with the different computational-basis
states of each subsystem, and thus quantify the scaling of the correlated SNR.
In the experiments discussed here, the subsystem measurement records have the
single-shot variances shown in Table.~\ref{tab:snr}. The corresponding SNR's
approach the cross-over between soft-averaging and thresholding.  While a
hybrid strategy could be envisaged, for simplicity we soft-average all
results. Then, the correlations computed using \eqref{eq:bare-corr} have
variance $\nu_{\mathrm{corr}}^2$ between 2.35 and 4.02 for the different
computational basis states. Thus, in order to resolve the correlated
measurement $\avg{\sigma_{z,1}\sigma_{z,2}}$ to the same accuracy as
$\avg{\sigma_{z,1}}$ we need $\approx 5$ times more averaging.

\begin{table}[tb]
  \begin{ruledtabular}
  \begin{tabular}{l|cccc}

  \textbf{State} & $\ket{00}$ & $\ket{01}$ & $\ket{10}$ & $\ket{11}$ \\
  \hline
    $M_1\; \nu^2$ & 0.42 & 0.44 & 0.85 & 0.77 \\
    $M_2\; \nu^2$ & 1.36 & 1.67 & 1.37 & 1.84 \\
    $\nu_\mathrm{corr}^2$ & 2.35 & 2.85 & 3.39 & 4.02 \\
  \end{tabular}
  \end{ruledtabular}
  \caption{\label{tab:snr}
  Variances of $M_1$ and $M_2$ measurement operators and correlation variance
  for various computational-basis states. Variance increases when the
  corresponding qubit is prepared in the $\ket{1}$ (excited) state, due to
  increased variance from relaxation events ($T_1$) during measurement. The
  corresponding single-shot readout fidelities are 0.59 and 0.18,
  respectively.}
\end{table}
  
\section{Tomographic Inversion}

Tomographic inversion is the process of converting a set of measurements into
a physical density matrix or process map. Since the estimates of single-
and multi-body terms may have unequal variances, the standard inversion
procedure should be modified to take these into account. The method below
provides an estimator for states and processes that may be readily fed into a
semidefinite program solver, if one wishes to add additional physical
constraints (such as positivity in the process map) \cite{Kosut2004}.

The correlated measurement records result in estimates of the 
expectation value $\langle \hat{M}_{(i,j,\cdots)}\rangle$
for the $N$-body observable
\begin{align}
\hat{M}_{(i,j,\cdots)} = 
\hat{U}_i^\dagger \hat M_0 \hat{U}_i \otimes
\hat{U}_j^\dagger \hat M_1 \hat{U}_j \otimes
\cdots.
\end{align}
If the state $\hat\rho$ of the system is initially unknown, the
problem of estimating $\hat\rho$ from a linearly-independent and
informationally complete set of observables $\hat{M}_{(i,j,\cdots)}$
can be cast as a standard linear regression problem. Similarly, if the
system evolves according to some unknown superoperator $\mathcal E$,
this superoperator can be estimated via linear regression by preparing
various known initial states $\hat\rho_\alpha$ and measuring a
linearly-independent and informationally complete set of observables
$\hat{M}_{(i,j,\cdots)}$. 

The best linear unbiased estimator for the reconstruction of a state
or process can be computed exactly for the cases where the covariance
matrix of the observations is known a priori---this corresponds to
generalized least-squares (GLS) regression. In the quantum case,
however, this covariance matrix depends on the state or process being
reconstructed, and so it is never known a priori---the best linear
unbiased estimator cannot be constructed. However, it can be
approximated by empirically computing the covariance matrix of the
observations. In the experiments described here, only the diagonal
elements of the covariance matrix were computed and used in the
reconstruction of the state or process, which leads to a slight bias
in the estimate, although the mean squared-error performance is still
good.

Formally, we can describe the state-tomography experiments as
follows. Let ${\rm vec}(\hat A)$ be the column-major vectorization of
an operator $\hat A$. Then the vector of state measurement expectation
values $m_s$ is given by
\begin{align}
m_s = P_{s} {\rm vec}(\hat\rho),
\end{align}
where
\begin{align}
P_{s} = \left[ 
\begin{array}{c}
{\rm vec}(\hat{M}_{(0,0,\cdots,0,0)})^\dagger\\
{\rm vec}(\hat{M}_{(0,0,\cdots,0,1)})^\dagger\\
\vdots
\end{array}
\right],
\end{align}
is the {\em state predictor matrix}, relating the state $\hat\rho$ to
the vector $m_s$ of expectation values. Then, if we have an estimate
$\tilde{m}_s$ for the expectation values, an estimator for $\hat\rho$ is
given by
\begin{align}
\tilde{\rho} = P_{s}^+ \tilde{m}_s
\end{align}
where
\begin{align}
P_{s}^+ = (P_s^\dagger C^{-1} P_s)^{-1} P_s^\dagger C^{-1},
\end{align}
is the GLS equivalent of the Moore-Penrose
pseudo-inverse of the state predictor, and where $C$ is the empirical
covariance for the measurements. In practice there are equivalent
alternatives to computing $\tilde{\rho}$ explicitly that have better
numerical stability properties.

Process-tomography experiments have an analogous description.  The main
distinction is that the predictor then maps a pair of input state and
measurement observable to expectation values, so the vector of expectation
values $m_p$ is given by
\begin{align}
m_p = P_{p} {\rm vec}(\mathcal E),
\end{align}
where $\mathcal E$ is the Liouville representation \cite{Blum2012} of the
process being characterized, and the {\em process tomography predictor
matrix} $P_{p}$ is given by
\begin{align}
P_{p} = \left[ 
\begin{array}{c}
{\rm vec}({\rm vec}(\hat{M}_{(0,0,\cdots,0,0)})^\dagger {\rm vec}(\hat\rho_{(0,0,\cdots,0,0)}))^\dagger\\
{\rm vec}({\rm vec}(\hat{M}_{(0,0,\cdots,0,1)})^\dagger {\rm vec}(\hat\rho_{(0,0,\cdots,0,0)}))^\dagger\\
\vdots\\
{\rm vec}({\rm vec}(\hat{M}_{(0,0,\cdots,0,0)})^\dagger {\rm vec}(\hat\rho_{(0,0,\cdots,0,1)}))^\dagger\\
{\rm vec}({\rm vec}(\hat{M}_{(0,0,\cdots,0,1)})^\dagger {\rm vec}(\hat\rho_{(0,0,\cdots,0,1)}))^\dagger\\
\vdots
\end{array}
\right],
\end{align}
where the input states $\hat\rho_{(i,j,\cdots)}$ are given by 
\begin{align}
\hat{\rho}_{(i,j,\cdots)} = 
\hat{U}_i \ketbra{0}{0} \hat{U}_i^\dagger \otimes
\hat{U}_j \ketbra{0}{0} \hat{U}_j^\dagger \otimes
\cdots.
\end{align}

\section{Tomography of Entangled States and Processes}

Given this machinery we can apply it to verfying the correlations in entangled
states and reconstructing processes that create entanglement. We use an
echoed cross-resonance interaction as an entangling two-qubit gate
\cite{Corcoles2013}. The single-qubit pulses used were 40 ns long and the total
duration of the refocused $\mathrm{ZX}_{-\pi/2}$ was 370 ns.

\begin{figure}
\includegraphics[width=\columnwidth]{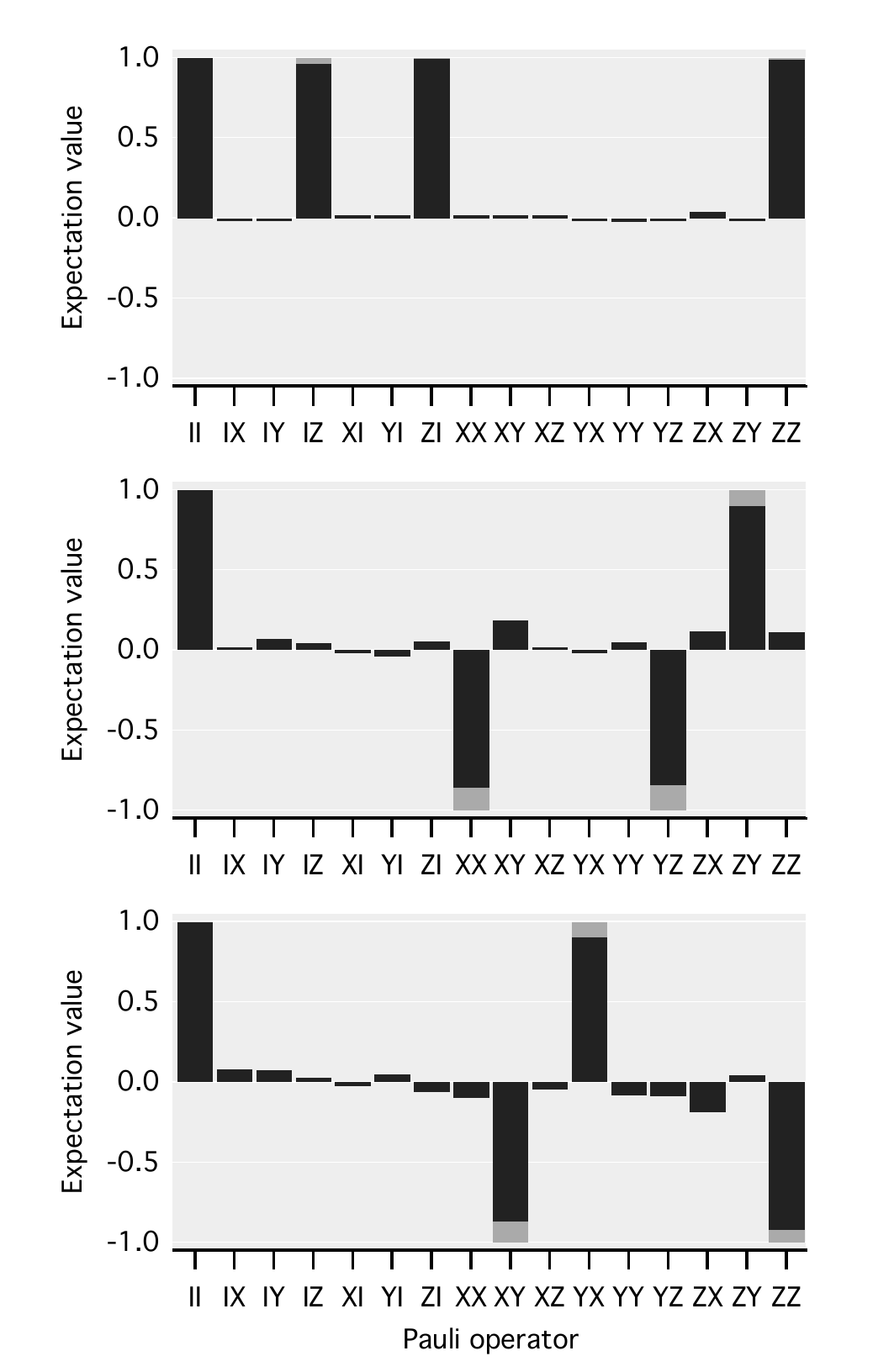}
\caption{
\label{TwoQubitPauliDecomps} Measured (black) and ideal (grey) Pauli
decompositions of three two qubit states showing both product and entangled
states. From top to bottom: the thermal ground state $\ket{00}$; the entangled
state after applying the $\mathrm{ZX}_{-\pi/2}$ gate to a $(I-Y)\otimes(I+Z)$
input state; the entangled state after applying $\mathrm{ZX}_{-\pi/2}$ to a
$(I-X)\otimes(I+Y)$ input. The fidelities, $F = \bra{\psi}\rho\ket{\psi}$, of
the entangled states to the ideal states $\ket{\psi}$ are 0.90 and 0.92,
respectively. The product state (top) is highlighted by the presence of weight-one
terms, whereas in the entangled states these terms are nearly zero.}
\end{figure}

Despite our imperfect single-shot readout we clearly witness high-fidelity
entanglement. The shot-by-shot correlation approach allows us to see
correlations that are not present in the product of the averages.  In
Fig.~\ref{TwoQubitPauliDecomps}(a) the product state shows a large response in
the individual measurements and two-qubit terms that are simply the product of
the single-qubit terms. Whereas in Fig.~\ref{TwoQubitPauliDecomps}(b and c), we
have an elegant demonstration of how maximally entangled states have only
correlated information: for the single-qubit operators, in all readouts we
observe a zero-mean response; however, certain two-qubit terms show maximal
response.

\begin{figure}
\includegraphics[width=\columnwidth]{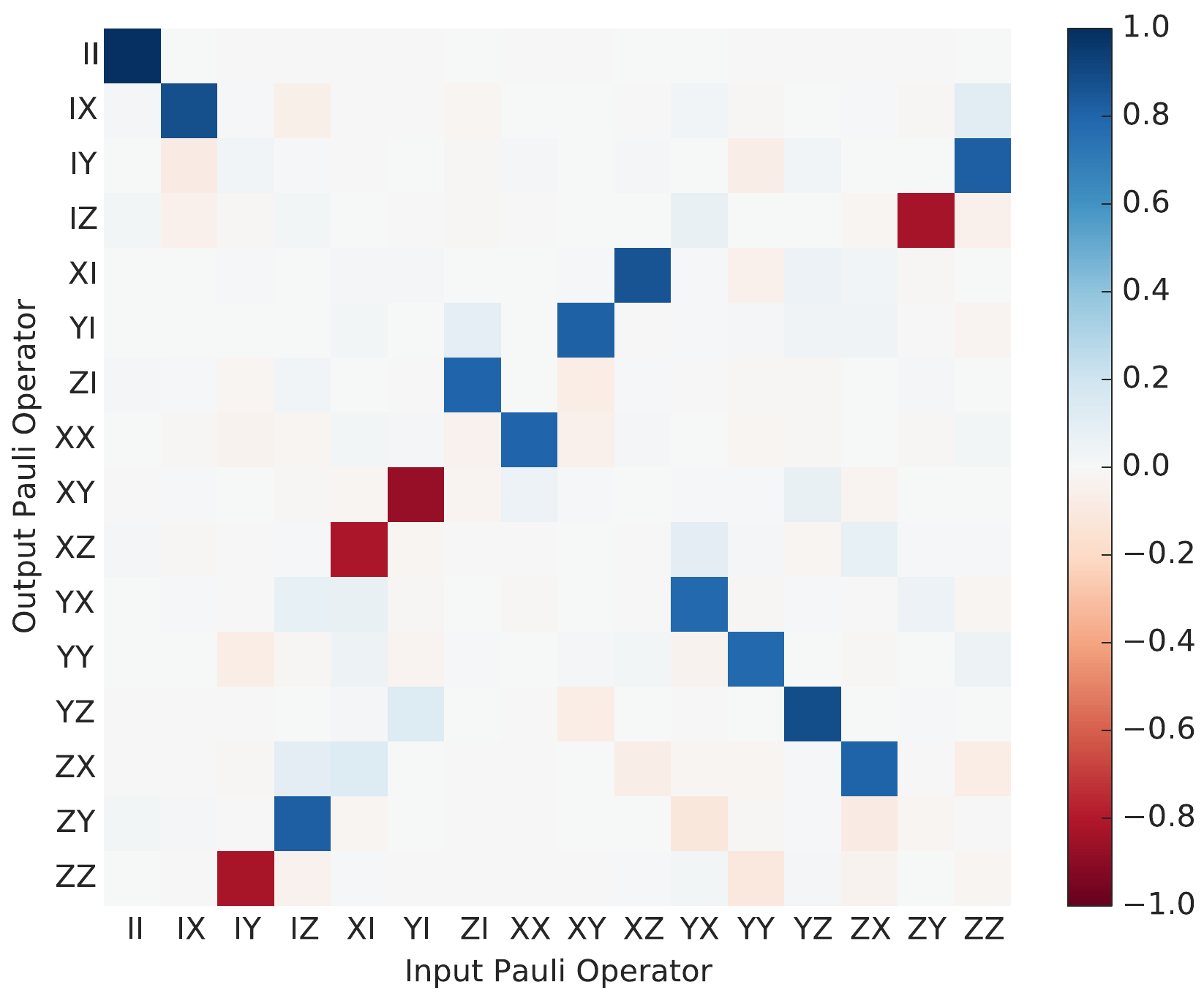}
\caption{\label{ProcessTomo} Pauli transfer matrix of a $\mathrm{ZX}_{-\pi/2}$ gate measured with correlated readouts.}
\end{figure}

Process tomography, shown in Fig.~\ref{ProcessTomo}, follows in a similar
fashion. Applying the procedure outlined above we find a gate fidelity for the
$\mathrm{ZX}_{-\pi/2}$ gate of 0.88. This process map clearly demonstrates
that our two-qubit interaction works on arbitrary input states, and that the
single-shot correlation strategy can recover information in arbitrary two-qubit
components in the resulting states.

\section{Conclusion}

Systems such as circuit QED with continuous measurment outcomes provide a
choice of strategies for qubit measurements.  We show that is is possible, and
sometimes preferrable, to directly correlate the continuous values without
thresholding.  This is an important tool when high-fidelity readout is not
available on all channels. We further have provided a straightforward protocol to
experimentally derive a nearly optimal linear filter to handle the transient
response of a pulsed measurement. Building on this we have constructed multi-qubit
measurement operators from shot-by-shot correlation of single-qubit
measurement records. The SNR of these correlated operators decreases
exponentially with the number of qubits, but in the high SNR regime multibody
correlations of a handful of qubits are still accessible. This provides a
framework for verifying quantum operations in architectures with imperfect
single-qubit measurements.

\begin{acknowledgments}

The authors would like to thank George A. Keefe and Mary B. Rothwell for
device fabrication. This research was funded by the Office of the Director of
National Intelligence (ODNI), Intelligence Advanced Research Projects Activity
(IARPA), through the Army Research Office contract no.~W911NF-10-1-0324. All
statements of fact, opinion or conclusions contained herein are those of the
authors and should not be construed as representing the official views or
policies of IARPA, the ODNI, or the U.S. Government.

\end{acknowledgments}

\bibliography{References} 

\end{document}